\documentclass[12pt,preprint]{aastex}
\usepackage{graphicx}
\usepackage{epstopdf}
\usepackage{natbib}
\pdfoutput=1

\begin{document}

\shortauthors{Prieto et~al.}

\title{Light Echoes from $\eta$~Carinae's Great Eruption: Spectrophotometric Evolution and the Rapid Formation
of Nitrogen-rich Molecules\altaffilmark{1}}

\author{J.~L.~Prieto\altaffilmark{2,3}, A.~Rest\altaffilmark{4}, F.~B.~Bianco\altaffilmark{5,6}, 
T. Matheson\altaffilmark{7}, N.~Smith\altaffilmark{8}, N.~R.~Walborn\altaffilmark{4}, 
E.~Y.~Hsiao\altaffilmark{9}, R.~Chornock\altaffilmark{10}, L.~Paredes~\'Alvarez\altaffilmark{11}, 
A.~Campillay\altaffilmark{9}, C.~Contreras\altaffilmark{9,11}, C.~Gonz\'alez\altaffilmark{9}, D.~James\altaffilmark{12}, 
G.~R.~Knapp\altaffilmark{2}, A.~Kunder\altaffilmark{13}, S.~Margheim\altaffilmark{14}, 
N.~Morrell\altaffilmark{9}, M.~M.~Phillips\altaffilmark{9}, R.~C.~Smith\altaffilmark{12}, 
D.~L.~Welch\altaffilmark{15}, A.~Zenteno\altaffilmark{12}}

\altaffiltext{1}{This paper includes data gathered with the 6.5 meter Magellan telescope 
at Las Campanas Observatory, Chile.}

\altaffiltext{2}{Department of Astrophysical Sciences, Princeton University, 4 Ivy Lane, Princeton, NJ 08544, USA} 

\altaffiltext{3}{Carnegie-Princeton Fellow}

\altaffiltext{4}{Space Telescope Science Institute, 3700 San Martin Dr., Baltimore, MD 21218, USA}

\altaffiltext{5}{Department of Physics, New York University, New York, N.Y., 10012}

\altaffiltext{6}{Center for Cosmology and Particle Physics, Department of Physics, New York University, 
New York, N.Y., 10012}

\altaffiltext{7}{National Optical Astronomy Observatory, Tucson, AZ 85719, USA}

\altaffiltext{8}{University of Arizona, Steward Observatory, Tucson, Arizona 85721, USA}

\altaffiltext{9}{Carnegie Observatories, Las Campanas Observatory, Casilla 601, La Serena, Chile}

\altaffiltext{10}{Harvard-Smithsonian Center for Astrophysics, 60 Garden Street, Cambridge, MA 02138, USA}

\altaffiltext{11}{Department of Physics and Astronomy, Aarhus University, Ny Munkegade 120, DK-8000 Aarhus C, Denmark}

\altaffiltext{12}{Cerro Tololo Inter-American Observatory, Casilla 603, La Serena, Chile}

\altaffiltext{13}{Leibniz-Institut f\"ur Astrophysik Potsdam, an der Sternwarte 16, 14482, Potsdam, Germany}

\altaffiltext{14}{Gemini Observatory, Southern Operations Center, Casilla 603, La Serena, Chile}

\altaffiltext{15}{Department of Physics and Astronomy, McMaster University, Hamilton, Ontario, L8S 4M1, Canada}

\begin{abstract}
We present follow-up optical imaging and spectroscopy of one of the
light echoes of $\eta$~Carinae's 19th-century Great Eruption
discovered by \cite{Rest12a}. By obtaining images and spectra at the
same light echo position between 2011 and 2014, we follow the
evolution of the Great Eruption on a three-year timescale. We find
remarkable changes in the photometric and spectroscopic evolution of
the echo light. The $i$-band light curve shows a decline of $\sim
0.9$~mag in $\sim 1$~year after the peak observed in early 2011 and a
flattening at later times. The spectra show a pure-absorption early
G-type stellar spectrum at peak, but a few months after peak the lines
of the \ion{Ca}{2} triplet develop strong P-Cygni profiles and we see
the appearance of [\ion{Ca}{2}]~7291,7324 doublet in emission. These
emission features and their evolution in time resemble those observed
in the spectra of some Type~IIn supernovae and supernova impostors.
Most surprisingly, starting \mbox{$\sim 300$~days} after peak
brightness, the spectra show strong molecular transitions of CN at
$\gtrsim 6800$~\AA. The appearance of these CN features can be
explained if the ejecta are strongly Nitrogen enhanced, as is observed
in modern spectroscopic studies of the bipolar Homunculus nebula.
Given the spectroscopic evolution of the light echo, velocities of the
main features, and detection of strong CN, we are likely seeing ejecta
that contributes directly to the Homunculus nebula.
\end{abstract}


\section{Introduction}

$\eta$~Carinae ($\eta$~Car) is one of the most well observed and, at
the same time, mysterious and poorly understood objects in the night
sky \citep[e.g.,][and references therein]{DHbook}. At a distance of $d
\simeq 2.35$~kpc \citep{Smith06} in the young cluster Trumpler~16 in
the Carina nebula, $\eta$~Car is a very luminous and massive evolved
star  ($L\sim 10^{6.7}$~L$_\odot$, $M\sim 100$~M$_\odot$; e.g.,
\citealt{Hillier2001}), a high mass-loss Luminous Blue Variable (LBV),
in an eccentric binary system \citep{Damineli96}. The system is
surrounded by a dusty, massive ($\sim 20$~M$_\odot$,
\citealt{Smith03}) bipolar nebula, the Homunculus nebula, that was
ejected $\sim 170$~years ago in the Great Eruption
\citep{Currie96,Smith98,Morse2001}.  

The Great Eruption (GE) of $\eta$~Car in the mid-1800s ($\sim
1840-1860$) was a spectacular astronomical event, a bright and
energetic transient visible to the naked eye that was observed and
registered by many \citep[e.g.,][]{Herschel1847}. Although we do not
understand the physical mechanism that caused $\eta$~Car's GE, it is
used as a standard reference for understanding episodic mass-loss in
very massive stars \citep[e.g.,][]{Smith14}, LBVs
\citep[e.g.,][]{Humphreys94}, supernova impostors
\citep[e.g.,][]{Humphreys99,vandyk02,Maund06,Smith11,Kochanek2012}, and the
most luminous supernovae \citep[e.g.,][]{Smith07}. Yet, until recently
we only knew it through visual estimates of its brightness and color,
due to the technological limitations at the time of this event
\citep{Frew04,SF11}. 

The recent discovery of $\eta$~Car's light echoes (LEs) by
\citet[][hereafter R12a]{Rest12a} now gives us the opportunity to
re-observe the multiple brightening events of $\eta$~Car in the
mid-1800s with modern instrumentation and from multiple directions
\citep{Rest12b}, similar to what has been done for ancient supernovae
in the LMC \citep{Rest05b,Rest08a}, SN~1987A \citep{Sinnott13}, Cas~A
\citep{Rest08b,Krause08a, Rest11_casaspec,Rest11_leprofile}, and Tycho
\citep{Rest08b,Krause08b}. In R12a we presented initial imaging and
spectroscopy of one of the brightest LEs of $\eta$~Car which showed
its unambiguous association with one of the brightenings during the GE
instead of the smaller eruption of circa 1890 \citep{Humphreys99}. The
spectra of this LE obtained in early 2011 showed pure absorption lines
consistent with supergiants with spectral type G2$-$G5 ($\rm T_{eff}
\sim 5000$~K) and line velocities of $\sim -200$~km~s$^{-1}$. 

In this letter, we present three years of imaging and spectroscopic
follow-up observations of the light echo of $\eta$~Car's GE discovered
by R12a. In Section~\S\ref{sec2} we discuss the observations and data
reduction. In Section~\S\ref{sec3} we present the results and analysis
of the light curve and spectra. We discuss our results in
Section~\S\ref{sec4}. 

\section{Observations}
\label{sec2}

\subsection{Imaging}

We observed the field of the $\eta$~Carinae LE from R12a at many
epochs between 2011 and 2014. Imaging was obtained with a variety of
telescopes and instruments, including: the MOSAIC~II and DECam
wide-field cameras mounted on the Blanco 4m telescope at Cerro Tololo
Inter-American Observatory (CTIO), the direct CCD camera mounted on the
Swope 1m telescope at Las Campanas Observatory (LCO), the Spectral CCD
camera mounted on the FTS 2m telescope at the Las Cumbres Observatory
Global Telescope Network (LCOGT) Siding Spring site \citep{Brown2013},
the SOAR Optical Imager mounted on the SOAR 4m telescope at the Cerro
Pach\'on, the Inamori-Magellan Areal Camera and Spectrograph
\citep[IMACS;][]{Dressler11}, and the LDSS~3 mounted on the
Magellan~Baade 6.5m telescope at LCO. Most of the images were obtained
using the SDSS $i$-band filter, with some epochs obtained using the
SDSS $grz$ filters, most of which were obtained with DECam.

\begin{figure}

\plotone{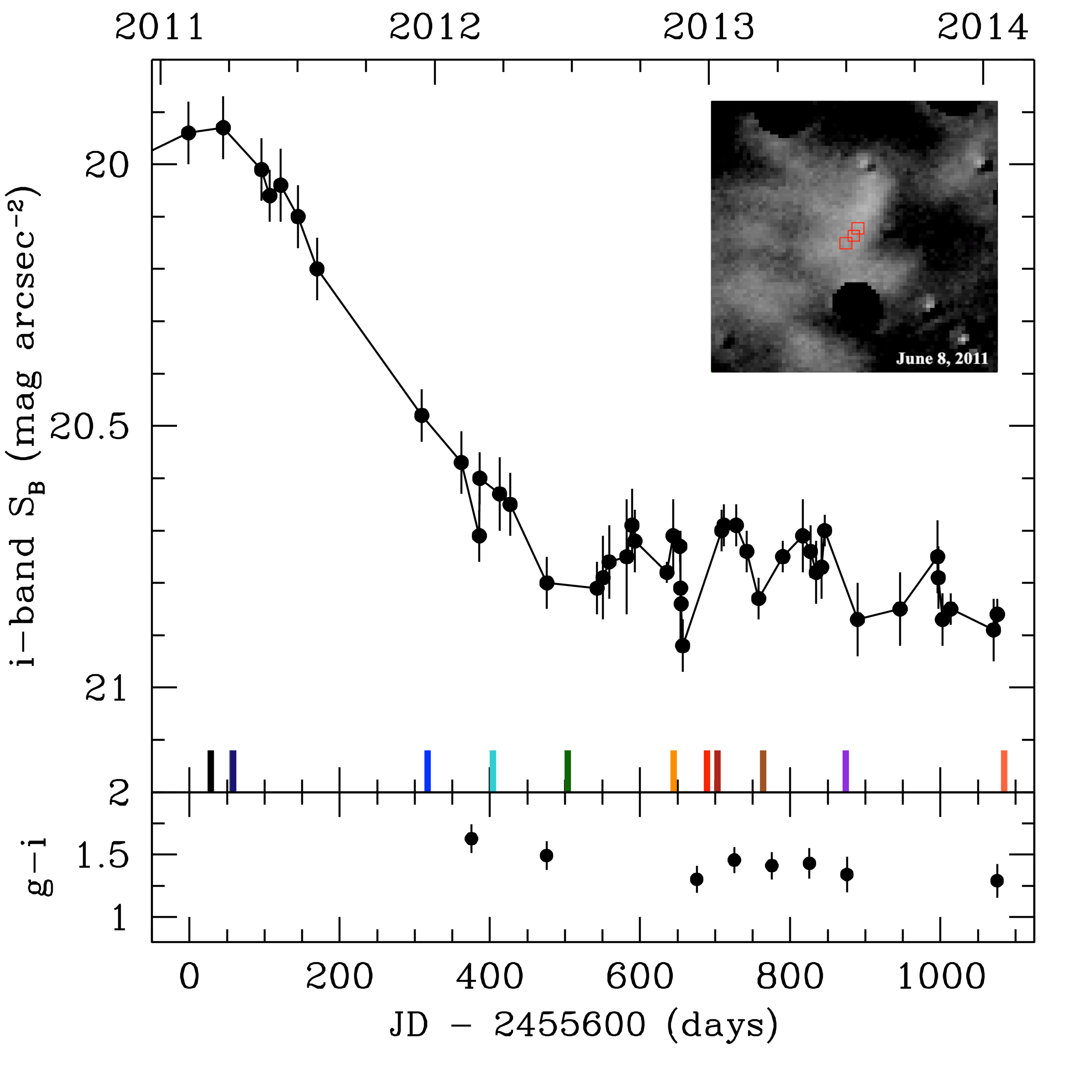}
\caption{Light curve of the light echo of $\eta$~Carinae's GE
discovered by R12a, obtained between early 2011 and 2014. The top
panel shows the $i$-band surface brightness and the bottom panel shows
the average $g-i$ color. This light curve and colors are obtained
doing photometry at the same echo position in all the images. The
inset shows $20\arcsec\times20\arcsec$ section of the difference
image from June 2011 centered on the light echo. The red boxes
are used to extract the light curve. The vertical lines show the times
at which optical spectra have been obtained.}
\label{fig1}
\end{figure}

Standard image reduction was performed on all the images, including
bias/overscan subtraction and flat-fielding using skyflats and
domeflats. We then ran a difference imaging analysis using the {\it
photpipe} pipeline \citep{Rest05a} that has been used to discover and
analyze the light echoes of historical supernovae
\citep[e.g.,][]{Rest05b,Rest08b} and $\eta$~Car (R12a). This pipeline
produces difference images that are very clean of point sources and
allow us to accurately measure the diffuse variable flux of the echo.
We used three $3 \times 3$ pixel ($0.8 \times 0.8\arcsec$) boxes
centered on the position of the LE (RA = 10:44:11.886, DEC =
$-$60:16:00.6, J2000.0), and along the position angle of the slit used for the
spectra (PA = 339 deg), to extract the difference flux light curve.
These fluxes were transformed into the DECam AB magnitude system using
observations of SDSS standards obtained in January 2014.
The light curve and mean color evolution are shown in
Figure~\ref{fig1}. A complete analysis of the imaging data and light
curves will be presented in Bianco et al. (2014, in prep.).  

\subsection{Spectroscopy}

We obtained several low-resolution ($R\sim 700-1700$) single-slit
optical spectra at the LE position between March~2011 and
January~2014. The spectra were obtained with several instruments,
including: IMACS mounted on the Magellan~Baade 6.5m telescope at LCO,
WFCCD mounted on the du~Pont 2.5m telescope at LCO, and the Gemini
Multi-Object Spectrograph (GMOS) mounted on the Gemini~South 8m
telescope at the Gemini Observatory (programs GS-2012B-Q-57,
GS-2013A-Q-11, GS-2013B-Q-19). Nod-and-shuffle techniques
\citep{nodandshuffle} were used with GMOS to improve sky subtraction. 

We reduced the IMACS and WFCCD spectra using the LA-Cosmic software
\citep{vandokkum2001} and standard routines in IRAF {\tt ccdred}, {\tt
twodspec} and {\tt onedspec} packages. The first IMACS spectrum was
reduced using custom IDL scripts (see R12a). The GMOS spectra were
reduced using the IRAF {\tt gemini} package and custom IDL scripts.
All the spectra were wavelength calibrated with an arc-lamp obtained
before or after the science exposures, and flux calibrated using a
spectrophotometric standard observed the same night. The 1D spectra
were extracted at a consistent position and with a spatial extraction
window of $\sim 10 \arcsec$ along the slit to increase the
signal-to-noise. We applied a telluric correction to all the spectra
that was extracted from the spectrophotometric standard star
observation. Figure~\ref{fig2} shows the spectral time sequence of
$\eta$~Car's LE.  

\section{Results and Analysis}
\label{sec3}

\subsection{Light Curve Evolution}

As we showed in R12a, the LE studied here is associated with
$\eta$~Car's GE because the total increase in magnitude observed
between 2003 and 2011 was $\gtrsim 2$~mag, which is inconsistent
with the lesser eruption of circa 1890. The light curve of
$\eta$~Car's LE presented in Figure~\ref{fig1} shows the observations
obtained between February 2011 and January 2014, a follow-up campaign
started after its discovery. The light curve has a peak between UT
February 6 and March 23 2011 at $i$-band surface brightness of
$19.9$~mag~arcsec$^{-2}$. After peak, it steadily declined by
$0.9$~mag in the first $\sim 450$~days. The mean $g-i$ color of the
light echo of $1.4$~mag does not change significantly during our
observations. 

At late times ($\gtrsim 450$~days past peak), after May 2012, there is
a significant flattening in the light curve and the LE stays at a
constant $i$-band surface brightness of $\sim 20.8$~mag~arcsec$^{-2}$
until our latest observations obtained in January 2014. In R12a we
concluded that this LE was most likely associated with the 1843
outburst observed in the historical light curve \citep[e.g.,][]{SF11}.
However, given the flat light curve at late times and absence of a
re-brightening thus far, which was observed in the historical light
curve in 1845, this now seems very unlikely. At present, the light
curve of this LE is more consistent with the 1838 peak or even the
1845 peak, but continuing photometric follow-up should give us a more
definite answer in the next few years. This LE light curve definitively
rules out the lesser eruption of circa 1890, which lasted $\sim
7$~years \citep[e.g.,][]{Humphreys99,SF11}. 

\subsection{Spectroscopic Evolution}

\begin{figure}
\plotone{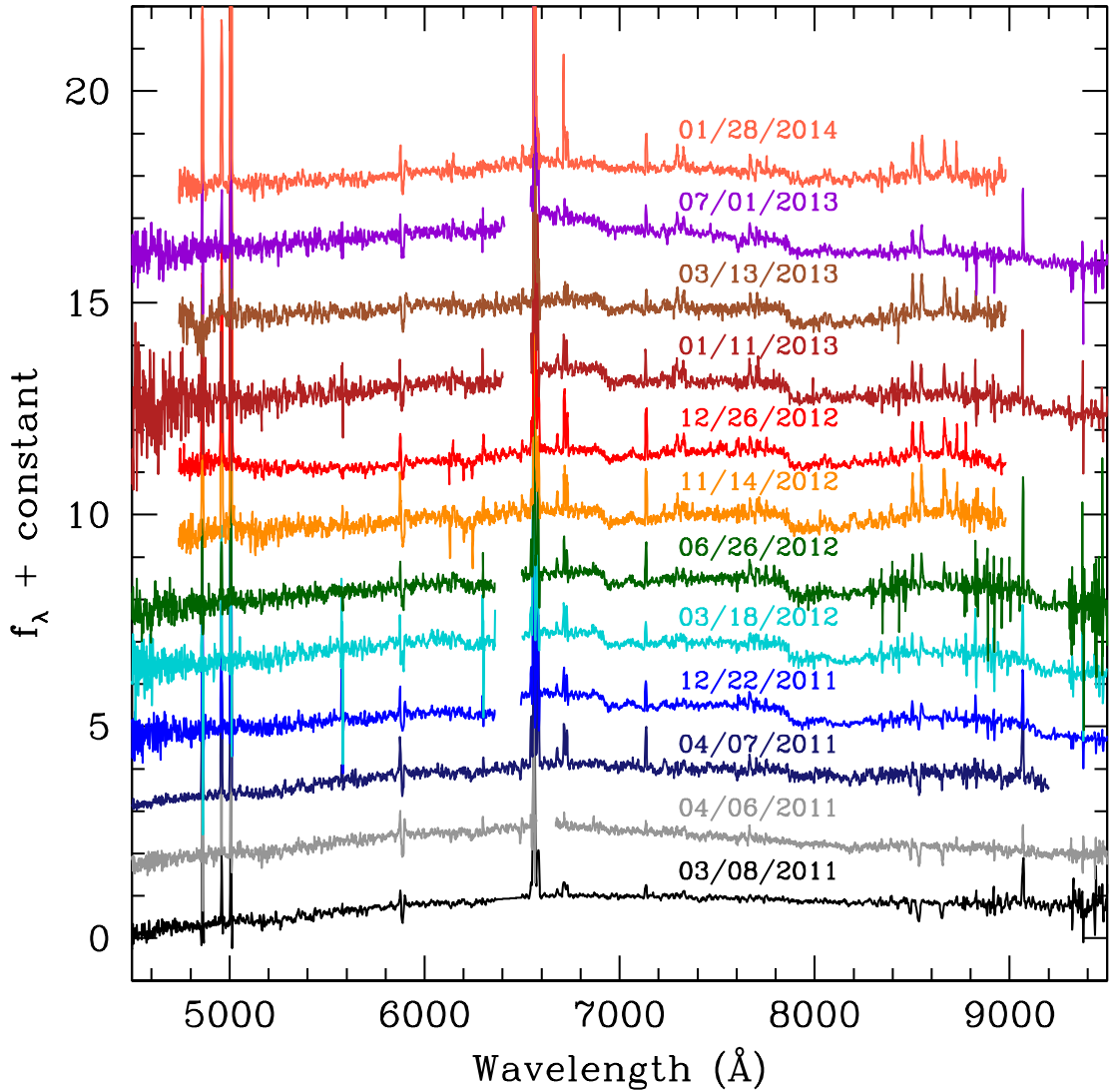}
\vspace*{-4cm}
\caption{Optical spectra of the light echo of $\eta$~Carinae's GE at different
epochs, from March 2011 (bottom) to January 2014 (top). Each spectrum has
been divided by a constant and we added a constant for clarity.}
\label{fig2}
\end{figure}

\begin{figure}
\plottwo{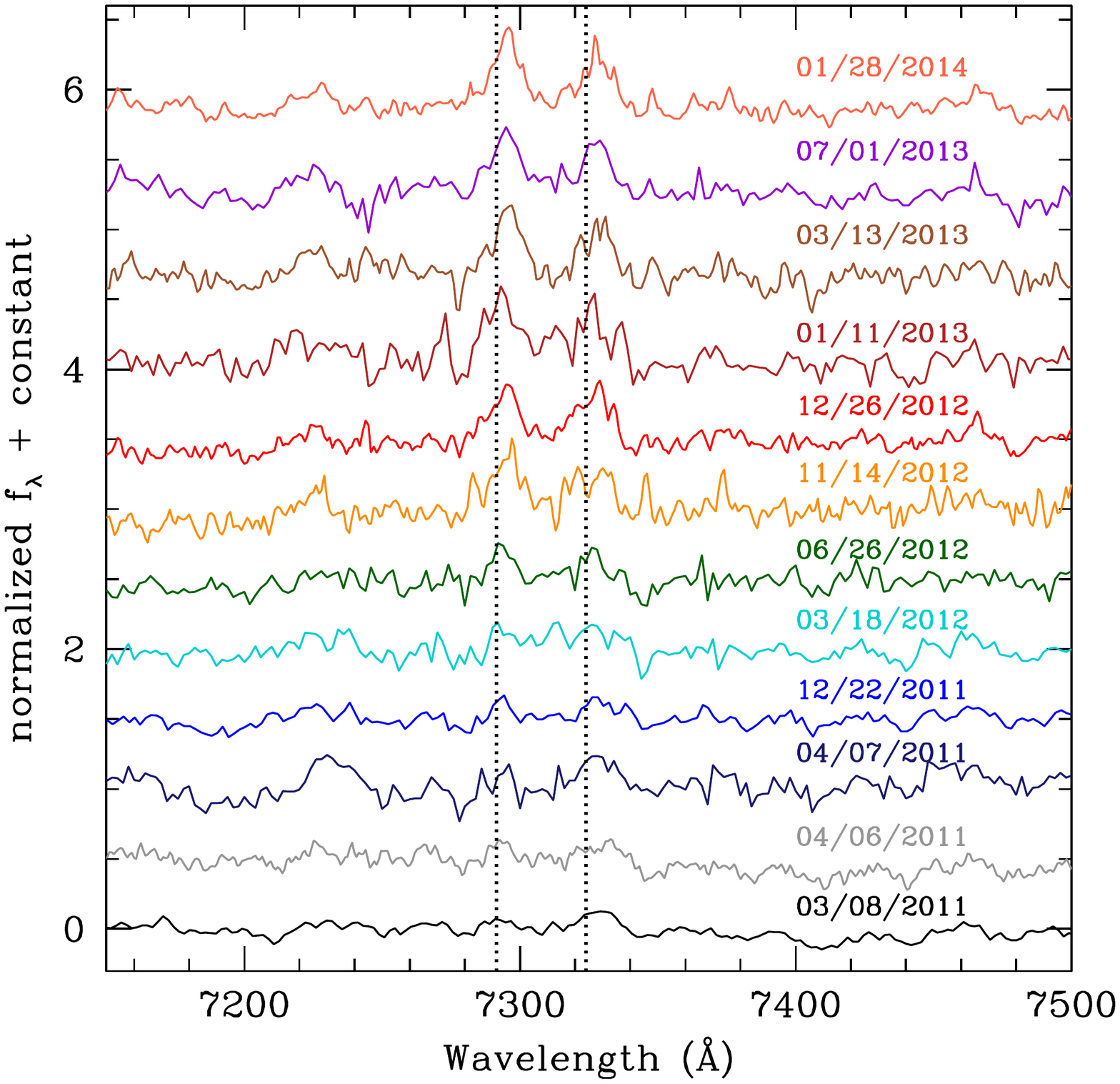}{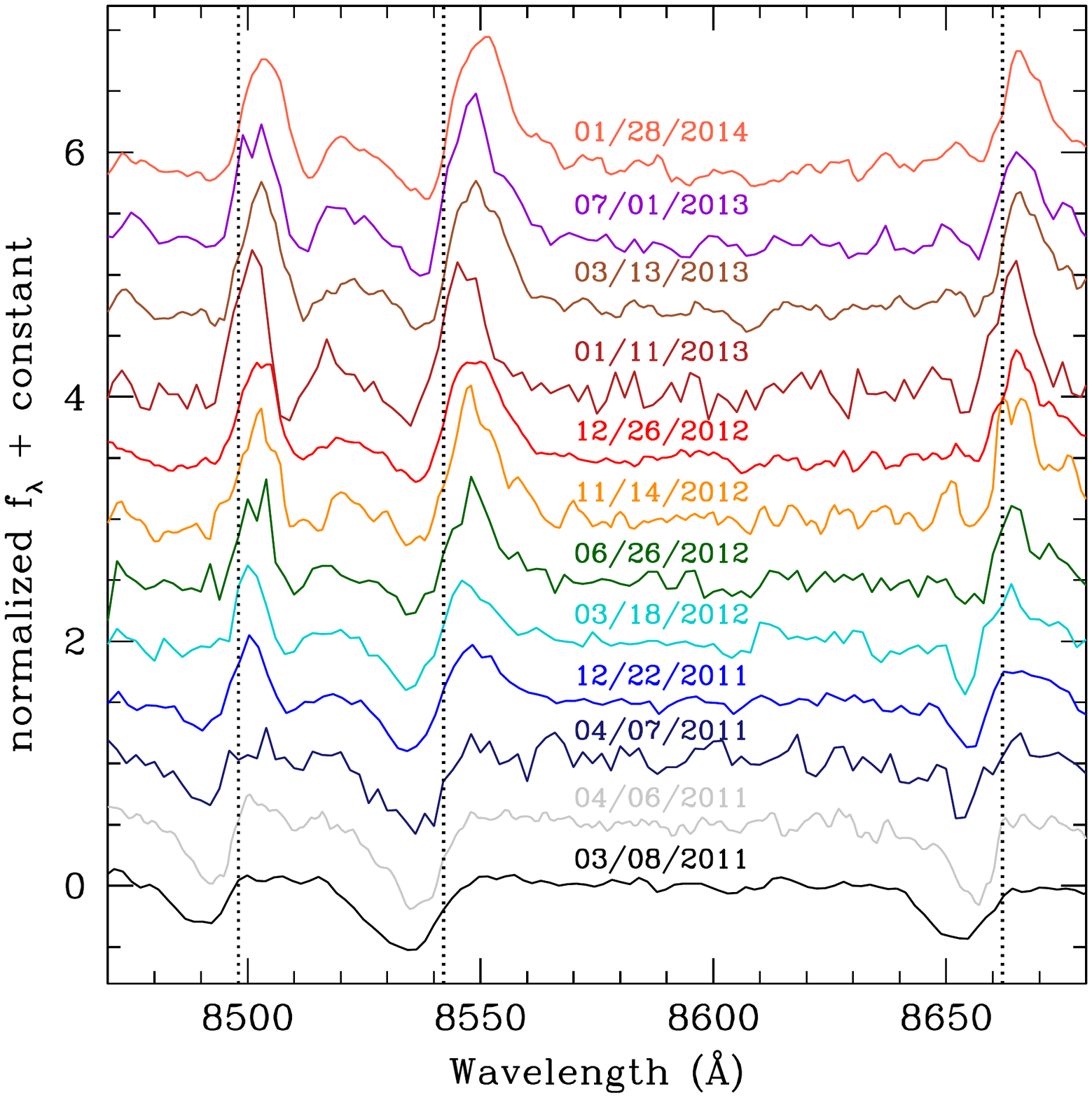}
\caption{Spectral evolution of the light echo in normalized flux versus
wavelength showing two different spectral regions centered on
\ion{Ca}{2} lines. {\bf Left:} Region of the spectrum around the
[\ion{Ca}{2}] 7291,7324 forbidden doublet (vertical lines). The emission 
lines are clearly detected at later times, starting March 2012. {\bf Right:} 
Region of the spectrum around the \ion{Ca}{2} 8498,8542,8662 triplet 
(vertical lines). The lines evolve from pure absorption features in 
March-April 2011 to P-Cygni profiles with absorption and emission starting 
in December 2011.}
\label{fig3}
\end{figure}

The evolution of the LE spectra is shown in Figures~\ref{fig2}
and \ref{fig3}. The first spectra of the echo obtained around peak
brightness in March-April 2011 are dominated by absorption
lines\footnote{All the strong emission lines observed in the spectra
(Balmer lines, [\ion{O}{3}], [\ion{N}{2}], [\ion{S}{2}], and
[\ion{Ar}{3}]) are from the surrounding nebula.} characteristic of
G2$-$G5 type supergiants (R12a). However, at later times (starting
December 2011, $\sim 300$~days after peak) the lines of the
\ion{Ca}{2} triplet have developed strong P-Cygni profiles (see
Figure~\ref{fig3}) with resolved absorption and emission components
($\rm FWHM \sim 400$~km~s$^{-1}$), characteristic of an expanding
photosphere. The emission line components of the \ion{Ca}{2} triplet
become stronger compared to the absorption at even later epochs. Also,
starting in the spectrum obtained on June 2012 ($\sim 500$~days after
peak) we see the appearance of the [\ion{Ca}{2}] 7291,7324 forbidden
doublet in emission. Despite the presence of the resolved \ion{Ca}{2}
lines in emission, we do not see a broad H$\alpha$ line component at
any time. We note that such a component could easily be masked by the
strong nebular lines of H$\alpha$ and [\ion{N}{2}] due to the low
resolution of our spectra.  

\begin{figure}
\plotone{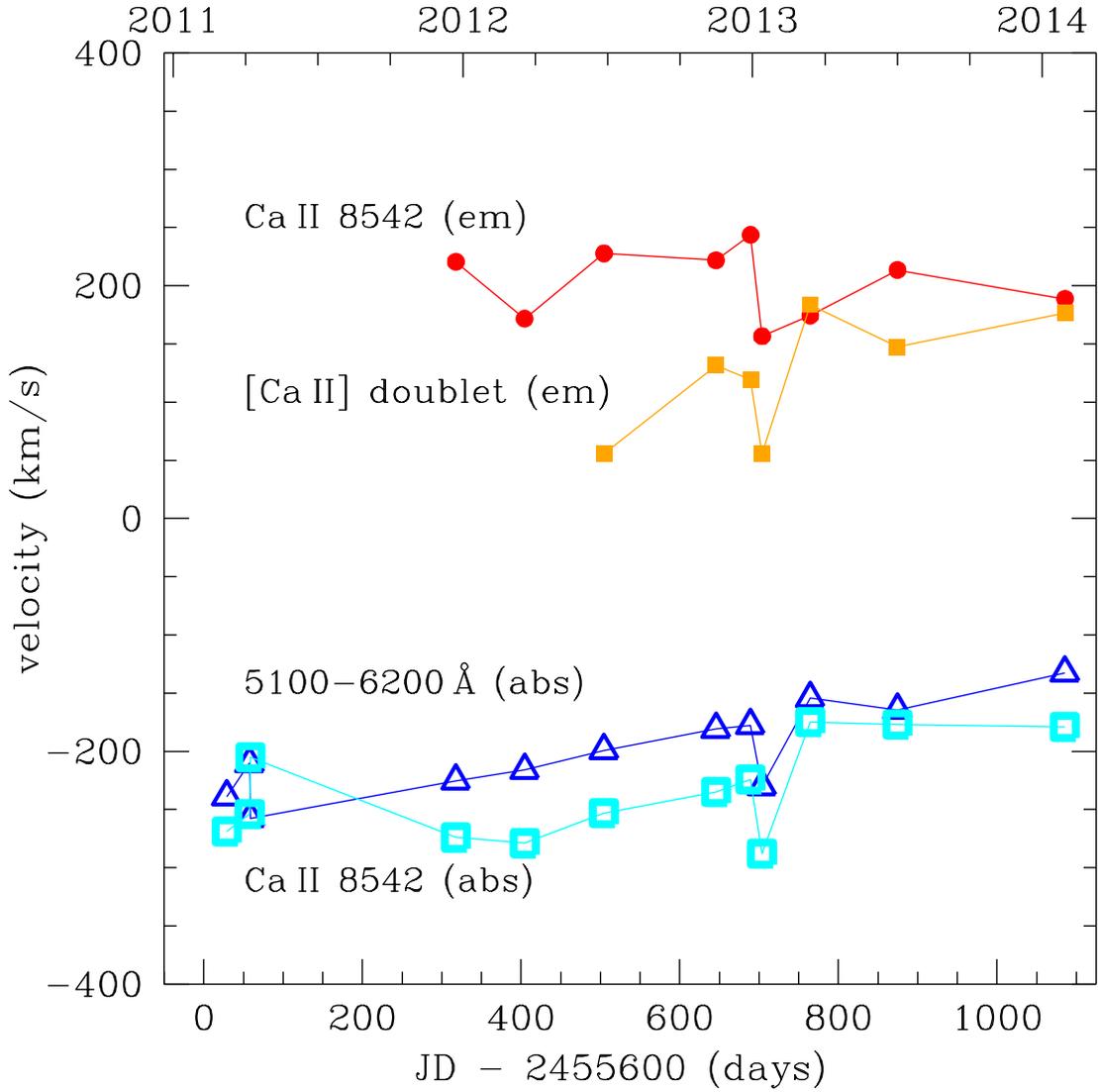}
\vspace*{-4cm}
\caption{Velocities of different features in the spectra as a function of 
time. The spectral features shown are: \ion{Ca}{2}~8542 in emission (filled circles), 
[\ion{Ca}{2}] 7291,7324 doublet in emission (filled squares), absorption 
lines in the $5100-6200$~\AA\ wavelength range (open triangles), and 
the minimum of the \ion{Ca}{2}~8542 absorption trough (open squares).}
\label{fig4}
\end{figure}

Figure~\ref{fig4} shows the velocities as a function of time of the
most significant features in the spectra of the LE. We used
cross-correlation techniques in IRAF with a set of stellar templates
and also the spectral fitting code
ULySS\footnote{\tt{http://ulyss.univ-lyon1.fr/}} \citep{Koleva2009} to
estimate the velocities of the absorption-dominated part of the
spectrum in the $5100-6200$~\AA\ region. Both techniques gave
consistent results, with mean velocity $v_{\rm abs} = -199 \pm
11$~km~s$^{-1}$, where the uncertainty is the error in the mean. For
the \ion{Ca}{2} emission lines, we measured the position of the peak
of the lines fitting a gaussian with {\tt  splot} in IRAF. The
measured mean velocity of the \ion{Ca}{2} 8542~\AA\ line is $v_{\rm
8542, em} = +202 \pm 9$~km~s$^{-1}$ and for the [\ion{Ca}{2}] doublet
is $v_{\rm [Ca\,II]} = +124 \pm 15$~km~s$^{-1}$. We also measure the
velocity of the minimum of the absorption trough of the \ion{Ca}{2}
8542~\AA\ line, which gives a mean velocity $v_{\rm 8542, abs} = -234
\pm 12$~km~s$^{-1}$. The shapes of the \ion{Ca}{2} triplet absorption
features in the first three spectra obtained March to April 2011 show
a clear blue asymmetry, with the blue edge of the absorption feature
extending to $\sim -850$~km~s$^{-1}$. The full line profiles at these
epochs can be well fitted with the addition of two gaussian components
with velocities of $-600$~km~s$^{-1}$ ($\rm EW \simeq 3$~\AA) and
$-200$~km~s$^{-1}$ ($\rm EW \simeq 7$~\AA). 

The defining characteristics of the late-time spectra of the LE are
the strong, broad absorption features seen in the $\gtrsim 6800$~\AA\
region (see Figure~\ref{fig2}). These strong features appear in the
spectrum obtained on December 2011 ($\sim 300$~days after peak) and
are present with similar strength in all the late-time spectra.
Figure~\ref{fig5} shows a zoom-in of the red part of the spectrum
obtained on December 2011, normalized at $\sim 6700$~\AA. The strong
absorption features observed in the light echo spectra match the
wavelengths of some of the rotation-vibration transitions
of the CN molecule. These molecular bands, which resemble steps in the
continuum between $\sim 7000-10000$~\AA, are usually observed in the
spectra of cool stars, such as carbon stars and red supergiants
\citep[e.g.,][]{WhiteWing1978}. 

The top spectrum in Figure~\ref{fig5} shows a red supergiant (RSG) 
model spectrum from \citet{Lancon2007} with $\rm T_{eff} =5000$~K, 
$\rm log(g) = -1.0$, and $\rm [Fe/H]= 0$. In order to approximately 
match the strengths of the CN steps observed in the LE
spectrum, we have multiplied the RSG model spectrum by a factor of 1.8 
after subtracting the continuum normalized at $\sim 6700$~\AA. 
The strongest molecular bands identified in the light echo 
spectra are the following transitions in the CN red system 
\citep{PearseGaydon1976}: ($v$',$v$'') = (3,0) at $6926$~\AA, 
(4,1) at $7089$~\AA, (2,0) at $7873$~\AA, 
(3,1) at $8067$~\AA, and (1,0) at $9141$~\AA. 

\section{Discussion}
\label{sec4}

We have presented three years of photometric and spectroscopic 
follow-up observations, from early 2011 to 2014, of the light echo of
$\eta$~Car's 19th century Great Eruption discovered by R12a. These 
observations give us a unique opportunity to study the evolution and 
physical properties of the GE in detail.  

The detection of P-Cygni profiles and the observed velocities of the
spectral features unambiguously link $\eta$~Car's GE to luminous
extragalactic transients associated with episodic mass-loss events in
massive stars, as has been suggested in previous studies based on its
historical light curve and the properties of the Homunculus nebula
\citep[e.g.,][]{Humphreys99,vandyk02,Smith11,Smith13}. Indeed, the
appearance and time evolution of strong \ion{Ca}{2} triplet, from pure
absorption features at peak to P-Cygni profiles to emission-dominated
profiles at late times, and [\ion{Ca}{2}] forbidden doublet in the
spectra of $\eta$~Car's light echo resemble the spectral properties
and evolution of some Type~IIn supernovae (SN~1994W,
\citealt{Dessart09}; SN~2011ht, \citealt{Mauerhan13}) and supernova
impostors \citep[e.g., UGC~2773-OT,][]{Smith10}. 

The time evolution of the line profiles of the LE is broadly
consistent with an optical depth effect in the ejecta, as seen in
Type~II SN. We can measure the electron density in the \ion{Ca}{2}
line-forming region using the ratio of intensities of the doublet to
the triplet \citep{FerlandPersson89}. From the light echo spectra, we
measure an intensity ratio of $\sim 1$, roughly constant at different
epochs. This gives an electron density in the range $\sim
10^8-10^{10}$~cm$^{-3}$ for $\rm T_e \simeq 3000-20000$~K. This high
density is consistent with modeling of the spectra of the Type~IIn
SN~1994W \citep{Dessart09}, which also shows a similar spectral
evolution in the \ion{Ca}{2} lines. However, we do not see in these
LE spectra the broad H$\alpha$ wings produced by electron
scattering in Type~IIn SN. The comparison to the SN impostor
UGC~2773-OT is particularly relevant here because the FWHM of the
H$\alpha$ line ($\sim 400$~km~s$^{-1}$) and its relative strength to
the \ion{Ca}{2} emission \citep{Smith10} are consistent with the
non-detection of broad H$\alpha$ features in $\eta$~Car's light echo
spectra. 

Despite the similarities between some of the properties of
$\eta$~Car's LE spectra and SN impostors, there are
important and interesting differences. As pointed out in R12a,
the LE spectrum at peak is consistent with cool supergiants
with late spectral type G2$-$G5 ($\rm T_{eff} \sim 5000$~K). This is
cooler than all the published spectral analysis studies of SN
impostors and LBV outbursts to date\footnote{For example, the coolest
well-measured temperature of an LBV in outburst is that of R71 in the
LMC, which had $\rm T_{eff} \sim 6650$~K in 2012 \citep{mehner13}.}.

The new, late-time spectra of the LE (starting in the spectrum
obtained $\sim 300$~days after peak) confirm this result. Spectral
fitting of the blue part of the spectra using UlySS and
cross-correlation with observed spectra of supergiants give $\rm
T_{eff}\sim 4000-4500$~K (Rest et al. 2014, in prep.). Furthermore, we
detect strong CN molecular bands in the red part of the optical
spectra, which are seen in the spectra of cool stars (e.g., carbon
stars and RSGs) but are not expected to be formed in warmer stellar
photospheres with $\rm T_{eff} \gtrsim 5500$~K
\citep[e.g.,][]{Lancon2007}. This indicates that the spectrum of the
LE likely becomes cooler after the light curve peak and stays cool at
late times. 

The cool temperature of the LE at peak and at late times after peak
are inconsistent with the most simple predictions from the opaque wind
model for the GE \citep{Davidson87}, and also very different from
S-Doradus variations and some LBV giant eruptions
\citep[e.g.,][]{Smith11}. The model of the GE as an explosion and
interaction with a strong pre-GE wind \citep{Smith13} appears to be
more consistent with our observations. However, we note that further
analysis and spectra of multiple light echoes as a function of time
will be needed in order to directly test different GE models 
\citep[e.g.,][]{Davidson87,Soker07,Smith13}. 

\begin{figure}
\plotone{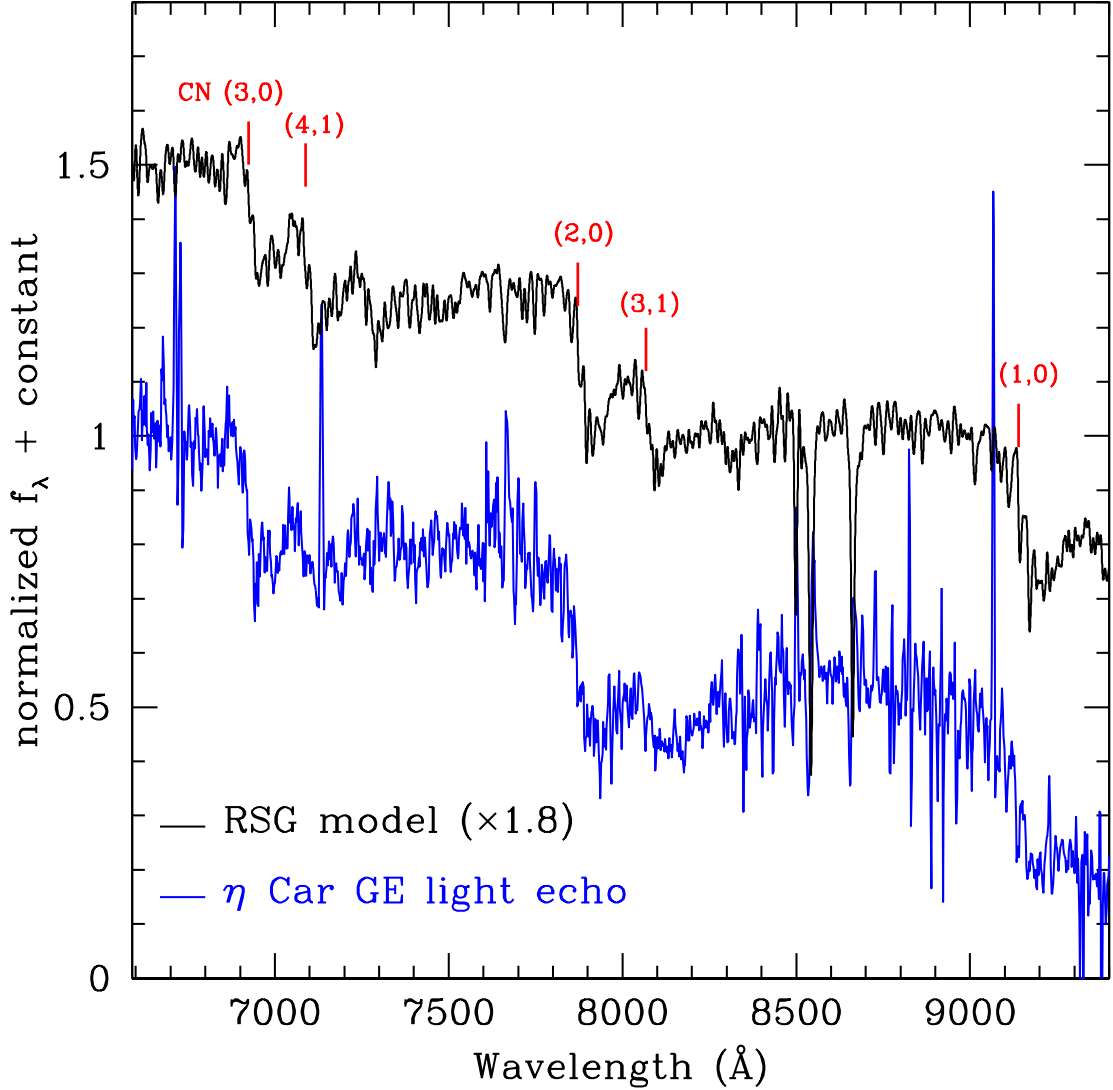}
\vspace*{-4cm}
\caption{Red section of the light echo spectrum from Dec. 22, 2011  (blue)
compared with the model spectrum of a RSG (black) with $\rm
T_{eff}=5000$~K and $\rm log(g) = -1.0$ from \citet{Lancon2007}. The
vertical lines show the wavelengths of the edges of the strongest
molecular bands of CN. The lowest transitions, particularly
(1,0), (2,0) and (3,0), are very strong in the light echo spectrum. In
order to approximately match the strengths of the CN steps in the
light echo spectrum, we have multiplied the spectrum of RSG model by 1.8
after its normalization at $\sim 6700$~\AA. The CN steps are present
with similar strength in all the light echo spectra obtained between
December 2011 and January 2014.}
\label{fig5}
\end{figure}

The CN features detected in the LE spectra at late times are
significantly stronger than expected from stellar atmosphere models
for RSGs (see Figure~\ref{fig5}). We could not find in the literature
any observed spectra with stronger features, relative to the
continuum, even among the $\sim 1000$ carbon stars identified in SDSS
(G.~Knapp, private communication) and also among different types of
massive star explosions and outbursts (SNe and SN impostors). This
points to an abundance effect, in the sense that the material ejected
in the GE is likely significantly rich in Nitrogen. This is consistent
with the measured over-abundance of N in the Homunculus nebula,
derived from ionized \citep[e.g.,][]{Davidson82,Dufour97} and from
molecular species \citep[e.g.,][]{Smith06b,Loinard12}. 

The spectroscopic evolution of the LE, the observed velocities of the
\ion{Ca}{2} features (which are consistent with the velocities of the
Homunculus at similar latitudes presented in \citealt{Smith06}), and
the strong CN features pointing to N enhancement, suggest that we are
seeing ejecta that contributes directly to the Homunculus nebula. 

\acknowledgments

We thank H.~Bond, J.~Dolence, J.~Murphy, and R.~Wing for discussions. 
We thank R.~Foley for help in one of the LCO runs.
We are indebted to the staff of Las Campanas, CTIO, LCOGT, and Gemini
observatories for their assistance. This work was supported by the HST
programs GO-12577, AR-12851, and GO-13486, and by NSF grant
AST-1312221. This work makes use of observations from the LCOGT
network. This paper is based in part on observations obtained at the
Gemini Observatory, which is operated by AURA, Inc., under a
cooperative agreement with the NSF on behalf of the Gemini
partnership: the NSF (USA), the NRC (Canada), CONICYT (Chile), the ARC
(Australia), CNPq (Brazil) and CONICET (Argentina). This project used
data obtained with the Dark Energy Camera (DECam), which was
constructed by the Dark Energy Survey (DES) collaborating
institutions. Funding for DES, including DECam, has been provided by
the U.S. DoE, NSF, MECD (Spain), STFC (UK), HEFCE (England), NCSA,
KICP, FINEP, FAPERJ, CNPq (Brazil), the GRF-sponsored cluster of
excellence ``Origin and Structure of the Universe" and the DES
collaborating institutions.

\newpage

%%%%%%%%%%%%%%%%%%%%%%%%%%%%%%%%%%%%%%

\end{document}